\documentclass[12pt]{article}

\usepackage{latexsym,amsfonts,amsmath,amssymb,pstricks,wasysym,stmaryrd,enumerate,epsfig}
\usepackage{mathrsfs}
\usepackage{theorem}

\numberwithin{equation}{section}

\newcommand{\be}{\begin{equation}}
\newcommand{\ee}{\end{equation}}
\newcommand{\beu}{\begin{equation*}}
\newcommand{\eeu}{\end{equation*}}
\newcommand{\bea}{\begin{eqnarray}}
\newcommand{\eea}{\end{eqnarray}}
\newcommand{\beaa}{\begin{eqnarray*}}
\newcommand{\eeaa}{\end{eqnarray*}}
\newcommand{\bmx}{\begin{pmatrix}}
\newcommand{\emx}{\end{pmatrix}}

\newcommand{\am}{\alpha}
\newcommand{\bm}{\beta}
\newcommand{\cm}{\gamma}

\newcommand{\as}{\mathsf a}
\newcommand{\bs}{\mathsf b}
\newcommand{\cs}{\mathsf c}
\newcommand{\ds}{\mathsf d}

\newcommand{\del}{\partial}

\newcommand{\Pp}{P_{{(+)}}}
\newcommand{\Pm}{P_{{(-)}}}
\newcommand{\Ppm}{P_{{(\pm)}}}
\newcommand{\Kin}{K^{-1}}
\newcommand{\tQ}{\mathcal Q}
\newcommand{\tS}{\mathcal S}
\newcommand{\mf}{\mathfrak}

\newcommand{\alf}{{\textstyle{\frac{1}{2}}}}
\newcommand{\half}{\frac{1}{2}}
\newcommand{\halfi}{\frac{i}{2}}

\newcommand{\quarter}{{{\frac{1}{4}}}}

\newcommand{\8}{{\infty}}

\newcommand{\liealg}[1]{{\mathfrak{#1}}}

\newcommand{\eps}{\epsilon}

\newcommand{\ket}[1]{{\,\left|#1\right>}\,}

\newcommand{\sqro}[1]{\left(#1\right)^\half}
\newcommand{\sqri}[1]{\left(#1\right)^{-\half}}
\newcommand{\rest}[1]{{#1}}
\newcommand{\mov}[1]{{#1}}

\advance\textheight by \topskip
\begin{document}

\baselineskip 17.5pt
\parindent 18pt
\parskip 8pt

\begin{flushright}
\break
arxiv:0704.2069 [hep-th]\\[3mm]
DCPT-07/09
\end{flushright}
\vspace{2cm}
\begin{center}
{\Large {\bf $q$-Deformed Supersymmetry and}}

{\Large {\bf Dynamic Magnon Representations}}\\[4mm]
\vspace{1cm} 
{\large C. A. S. Young}
\\
\vspace{0.4cm}
{\em Department of Mathematical Sciences, University of Durham,\\
South Road, Durham DH1 3LE, UK}
\\
{\small E-mail: {\tt charles.young@durham.ac.uk}}
\\

\end{center}

\vskip 0.2in
 \centerline{\small\bf ABSTRACT}
\centerline{
\parbox[t]{5in}{\small
\noindent
It was recently noted that the dispersion relation for the magnons of planar $\mathcal N=4$ SYM can be identified with the Casimir of a certain deformation of the Poincar\'e algebra, in which the energy and momentum operators are supplemented by a boost generator $J$. By considering the relationship between $J$ and $\liealg{su}(2|2)\ltimes \mathbb R^2$, we derive a $q$-deformed super-Poincar\'e symmetry algebra of the kinematics. Using this, we show that the dynamic magnon representations may be obtained by boosting from a fixed rest-frame representation. We comment on aspects of the coalgebra structure and some implications for the question of boost-covariance of the $S$-matrix.
}}

\vspace{1cm}

\newpage
\section{Introduction}
In the study of anomalous dimensions in planar $\mathcal N=4$ supersymmetric Yang-Mills \cite{MZ,KMMZ,Bth,PlefRev,BS}, it has proven very profitable to pick out a preferred $R$-symmetry generator $L$ and then focus on states whose charge under $L$ and conformal dimension $\Delta$ are both large, but with $\Delta-L$ finite \cite{BMN,Staudacher04}. Single-trace operators in this sector may be regarded as long spin-chains, most of whose sites are in the ``vacuum'' state $Z$ ($\Delta=L=1$), with the finite number of other SYM fields in the trace regarded as particle-like excitations, called magnons. One can set up a scattering theory of these magnons in which the dynamics are governed by a factorizable $S$-matrix \cite{Staudacher04}. Given the $S$-matrix, the energy (i.e. dilatation) spectrum can in principle be computed via algebraic bethe ansatz techniques.

The $S$-matrix for the complete set of elementary magnons was first constructed by Beisert in \cite{BeisertSU22}, and is determined by symmetry considerations up to one overall ``dressing'' factor $S_0(p_1,p_2,g)$, a function of the magnon momenta and the coupling. The dressing factor, as a means of interpolation from weak to strong coupling, was first introduced in \cite{AFS}, where its general structure in terms of certain conserved charges was also conjectured.

In the usual relativistic scattering theories in 1+1 dimensions (see e.g. \cite{Dorey} for a review), there are well-established physical conditions that $S$-matrices should obey and which are used to constrain such factors. These include symmetry under crossing (exchange of an in- with an out-channel) and the bootstrap principle (which relates simple poles in the analytically continued $S$-matrix to bound states of the model).  It was shown by Janik \cite{Janik} that there is a natural analogue of the crossing relation for the non-relativistic magnon $S$-matrix, and
that if the $S$-matrix is to obey this relation then the dressing factor must satisfy an additional equation. Drawing on the results of \cite{AFS,HL,AF06}, dressing factors obeying Janik's equation were proposed in \cite{BES,BHL}, and recently the pole structure of the $S$-matrix with these factors was shown \cite{DHM} to be compatible with the known spectrum of BPS magnon bound states \cite{NDorey,CDO}. See \cite{r} for further recent progress. 

The remaining physical condition from relativistic exact $S$-matrix theory, which has as yet found no role in the case at hand, is in many ways the simplest: Lorentz covariance itself. Given that this is such a powerful constraint (forcing relativistic $S$-matrices to depend on the particles' momenta only through the difference of their rapidities) it is important to establish whether there is any analogous symmetry here.

Some evidence that there might be was uncovered in \cite{GH07}, which identified a generator of ``deformed boosts'' acting on the elliptic rapidity plane. This is reviewed in section \ref{Eq11}. In section \ref{algebra} we discuss the behaviour of the supercharges $Q$ and $S$ under the boost generator. We argue that $Q$ and $S$ should be viewed as generators of a $q$-deformed super-Poincar\'e algebra, which we then go on to construct in detail. This is the main result of the paper; as a consequence, we show that the dynamic representations may be obtained by finite boosts from a rest-frame representation. The classical limit of the deformed supersymmetry algebra in the plane-wave regime is also obtained. Finally, in section \ref{coalgebra}, we comment on the coalgebra structure and make some remarks on what boost-covariance of the $S$-matrix would mean in this deformed setting. In particular, we argue that, in contrast to the usual un-deformed case,  knowledge of the full supersymmetry algebra is a prerequisite for determining whether the system is boost-covariant.

\section{Boosts and the uniformizing variable}\label{Eq11}
We begin by recalling briefly the argument of \cite{GH07}. In relativistic quantum mechanics in 1+1 dimensions, particles transform in irreducible representations of the Poincar\'e algebra $E(1,1)$, 
\be [J,P]=E,\quad [J,E]=P,\quad [E,P] = 0,\ee
where $J, E, P$ are the generators of, respectively, Lorentz boosts and time- and space-translations. An irreducible representation is selected by specifying a value for the Casimir
\be m^2 = \mathscr{C} = E^2 - P^2,\ee
and this equation is the usual relativistic dispersion relation.

It is natural to ask whether there is a similar interpretation, as the Casimir of some algebra of kinematical symmetries, for the magnon dispersion relation
\be \frac{1}{4} = C^2 - 4g^2 \sin^2 \left(\frac{P}{2}\right) \label{disprel}.\ee 
Here $C$ is the $\liealg{su}(2|2)$ central charge, $P$ is the magnon momentum, and $g^2$ is proportional to the t'Hooft coupling \cite{BDS}. As was observed in \cite{GH07}, there exists a deformation of the Poincar\'e algebra, denoted $E_q(1,1)$, whose Casimir has the correct form. It was introduced in \cite{CGST} and is defined to be the unital algebra generated by $E$, $K$ and $J$, subject to the relations 
\be K E = E K, \quad K K^{-1} = 1\label{dr}\ee 
\be K J = J K - aiEK,\quad[J,E] = \frac{1}{2ai} \left( K - K^{-1} \right).\label{dr2}\ee 
Here $a$ is a real number related to the deformation parameter by $q=e^{ia}$. (If one writes $K=e^{ia\tilde P}$ then in the limit $a\rightarrow 0$ one recovers the usual Poincar\'e algebra with generators $E, \tilde P $ and $J$.)  The Casimir is 
\be \mathscr C = a^2 E^2 + K + K^{-1} - 2 \label{cas}\ee 
and this is equivalent to the dispersion relation (\ref{disprel}) provided we make the identifications 
\be C= ag E \qquad K = e^{iP}\label{id1}\ee 
and set 
\be \mathscr C = \frac{1}{4g^2} \, \label{casval}.\ee
Note that one can consistently interpret $a$ and $E$ as having, respectively, dimensions of length and inverse length, and $J$ and $K$ as being dimensionless.

As in the usual relativistic case, when we consider representations in which the two commuting generators take definite values $E$ and $K$ then the identification of the boost generator provides a systematic way of introducing the uniformizing parameter $z$  \cite{Janik} on the space of on-shell pairs $(E,K)$. One demands that $J$ be realized as $\del/\del z$, and then the algebraic relations (\ref{dr2}) yield differential equations for $E(z)$ and $P(z)$, which may be solved in terms of elliptic functions \cite{Janik,Beisert06,GH07}.

\section{The deformed supersymmetry algebra}\label{algebra}
At this stage what we have is  the generator $J$ of infinitesimal translations in the generalized rapidity plane. We now discuss how $J$ is related to the other symmetries of the kinematics. 

The excitations of the scattering theory transform in representations of the centrally extended superalgebra\footnote{Or rather the product of two copies of this algebra, with the central charges identified. But it is possible, and simplifies matters, to focus on only one copy.} $\liealg{su}(2|2) \ltimes \mathbb{R}^2$ (we summarize the relevant facts below; for full details see \cite{BeisertSU22,Beisert06}) and one of the striking things is that these representations are ``dynamic'', i.e. dependent on the momentum $P$ of the excitation. More precisely, for a given multiplet -- for example, the fundamental multiplet $\{\ket{\phi^a},\ket{\psi^\am}\}$ of elementary magnons -- the action of the even-graded $\liealg{su}(2)\times \liealg{su}(2)$ generators is fixed, but the action of the odd-graded generators $Q^\am_a$ and $S^a_\am$ is momentum-dependent. 

This is certainly odd if one thinks of $\liealg{su}(2|2) \ltimes \mathbb{R}^2$ as an algebra of \emph{internal} symmetries. The point of view we take here is that it is more natural to regard $Q^\am_a$ and $S^a_\am$ as \emph{spacetime} supersymmetries of the 1+1 dimensional scattering theory. Indeed, since the action of $J$ on any state alters the value of $K=e^{iP}$, according to (\ref{dr2}), it follows that the dynamic generators $Q^\am_a$ and $S^a_\am$ cannot commute with $J$ -- that is, they cannot be singlets of the deformed Poincar\'e algebra. 
One should therefore ask: what are the algebraic relations between $J$ and the $Q^\am_a$ and $S^a_\am$? It is this question we address now. With the familiar super-Poincar\'e algebra in mind, we expect that the supersymmetries transform as ``deformed'' spinors, and we shall see that this idea can be made precise.

Recall that the even generators of the superalgebra $\liealg{su}(2|2) \ltimes \mathbb{R}^2$ are $L^\am{}_\bm$ and $R^a{}_b$, of $\liealg{su}(2) \times \liealg{su}(2)$, together with central charges $C, \Pp, \Pm$.
The odd generators $Q^\am_a$, $S^a_\am$ transform canonically under $\liealg{su}(2) \times \liealg{su}(2)$,
\be [L^\am{}_\bm, Q^\cm_c ] =   \delta^\cm_\bm \, Q^\am_c - \alf \delta^\am_\bm \, Q^\cm_c  \qquad [R^a{}_b, Q^\cm_c] = -\delta^a_c \, Q^\cm_b + \alf\delta^a_b \, Q^\cm_c\ee
\be [L^\am{}_\bm, S^c_\cm ] = - \delta^\am_\cm \, S^c_\bm +\alf \delta^\am_\bm \, S^c_\cm  \qquad [R^a{}_b, S^c_\cm] =   \delta^c_b \, S^b_\cm-\alf \delta^a_b \, S^c_\cm, \ee
and close into the even subalgebra according to
\bea \{ Q^\am_a , Q^\bm_b \} &=& \eps^{\am\bm} \eps_{ab} \Pp \label{fc}\\
     \{ S^a_\am , S^b_\bm \} &=& \eps^{ab}\eps_{\am\bm} \Pm \\
      \{ Q^\am_a,S^b_\bm \} 
  &=& \delta^b_a L^\am{}_\bm + \delta^\am_\bm R^b{}_a 
  + \delta^b_a \delta^\am_\bm C \, .\eea
It turns out to be necessary to relate the values of the charges $\Ppm$ to the momentum $P$ according to\footnote{We suppress the extra degree of freedom $\alpha$ of \cite{Beisert06} for simplicity.}
\be \Pp = g(1 - K),\qquad  \Pm = g(1-\Kin).\label{PandK}\ee
A number of reasons for this were given in \cite{BeisertSU22,GH06,PST,Beisert06}, differing in emphasis but all closely related to the form of the coproduct $\Delta\Ppm$. We would like to concentrate in this section on the algebra structure and single-excitation states, reserving discussion of the coalgebra structure and multiple excitations for section \ref{coalgebra}, but it is helpful to introduce $\Delta\Ppm$ here briefly. The $\Ppm$ are in origin operators which, for each excitation of the chain in turn, add/remove a vacuum site $Z$ next to that excitation. On the asymptotic scattering states with which we are concerned, this causes other excitations, although by assumption distant from the point at which $\Ppm$ acts, to pick up phases, and this non-locality is encoded in the comultiplication rules \cite{GH06,PST}
\bea \Delta \Pp &=& \Pp \otimes K + 1 \otimes \Pp \label{DPpm}\\
     \Delta \Pm &=& \Pm \otimes \Kin + 1 \otimes \Pm \eea
which specify how $\Ppm$ act on tensor products, i.e. on multi-particle states.
For our purposes the important point is that it is consistent \cite{PST} to make the identifications (\ref{PandK}) at the algebraic level: they are clearly compatible with the algebra structure, since $K$ and $\Ppm$ are central (this step precedes the introduction of $J$); for the coalgebra structure, one may easily check (\ref{PandK}) are compatible with (\ref{DPpm}) together with the usual additivity property of momenta, $\Delta P = 1 \otimes P + P \otimes 1$ $\Rightarrow$ $\Delta K= K \otimes K$. Thus we may replace the $\Ppm$ everywhere they occur and cease to treat them as independent generators.

\subsubsection*{Derivation of the algebra relations} After making the identifications (\ref{id1}) and (\ref{PandK}), the complete set of even generators is $L^\am{}_\bm,R^a{}_b,E,K$ and $J$, and the closure relations (\ref{fc}) of the odd generators become
\bea \{ Q^\am_a , Q^\bm_b \} &=& g\eps^{\am\bm} \eps_{ab} (1-K) \label{QQ}\\
     \{ S^a_\am , S^b_\bm \} &=& g\eps^{ab}\eps_{\am\bm} (1-\Kin) \label{SS}\\
      \{ Q^\am_a,S^b_\bm \} 
  &=& \delta^b_a L^\am{}_\bm + \delta^\am_\bm R^b{}_a 
  + \delta^b_a \delta^\am_\bm\,ag E\, .\label{QS}\eea
Our goal is to extend the algebra $E_q(1,1)$ defined in (\ref{dr2}).
We may take the $L^\am{}_\bm$ and $R^a{}_b$ to be internal symmetries, commuting with $J$, since their action on states is known to be $P$-independent. The most general form of bracket of $J$ with $Q^\am_a$ and $S^a_\am$ compatible with this $\liealg{su}(2) \times \liealg{su}(2)$ structure and linear in the supersymmetries is
\be [J,Q^\am_a] = A Q^\am_a + B \eps^{\am\bm} \eps_{ab} S^b_\bm \ee
\be [J,S^a_\am] = C S^a_\am + D\eps^{ab} \eps_{\am\bm} Q^\bm_b \ee
where $A,B,C,D$ are $\liealg{su}(2) \times \liealg{su}(2)$ singlets (and graded even) but are otherwise unknown.

The commutator of $J$ with  $\{Q^\am_a, Q^\bm_b\}$ may then be computed in two ways: on the one hand
\be \left[J, \{Q^\am_a, Q^\bm_b\}\right] = \eps^{\am\bm}\eps_{ab} g [J,1-K] =
                    - \eps^{\am\bm}\eps_{ab} g[J,K] = -iag \eps^{\am\bm}\eps_{ab} EK  \ee
using (\ref{dr2}), but at the same time, using the graded Jacobi identity, the relations (\ref{QQ}-\ref{QS}) and the invariance of the $\eps$ symbol,
\be \left[J, \{ Q^\am_a, Q^\bm_b\}\right] =  \left\{ [J,Q^\am_a], Q^\bm_b\right\} + \left\{ [J,Q^\bm_b], Q^\am_a\right\} = 2g \eps^{\am\bm}\eps_{ab} \left( A(1-K) + aBE \right) .\ee
(We assume here that $A,B$ commute with $Q^\am_a$).
There is therefore a constraint
\be -iaEK = 2\left( A(1-K) + aBE \right).\label{c1}\ee
Similarly, it follows from consideration of $\{ S^a_\am,S^b_\bm\}$ that
\be ia E \Kin = 2\left( C(1-\Kin) + a DE \right)\label{c2}\ee
and from the remaining bracket, $\{Q^\am_a,S^b_\bm\}$, that \be A+C=0\ee and
\be -\frac{ig}{2}(K-\Kin) = g(1-\Kin) B + g(1-K) D.\label{c3}\ee 
Equations (\ref{c1}) to (\ref{c3}) are solved by
\bea A= -\halfi \lambda aE,&\quad& B= \halfi \lambda - \halfi (\lambda+1) K ,\\
    C =\halfi \lambda aE, &\quad& D =-\halfi \lambda + \halfi (\lambda+1) \Kin,\eea
with $\lambda\in\mathbb C$, yielding a one-parameter family of allowed brackets of $J$ with the supersymmetries.

It is clear that setting $\lambda=0$ simplifies the brackets somewhat, but in fact on inspection there turns out to be another choice which makes the relations more symmetrical and which will be useful in the following sections. We let $\lambda=-\frac{1}{2}$, so that
\bea  \left[J,Q^\am_a\right] &=&+\frac{i}{4} a E Q^\am_a - \frac{i}{4} \left(1+K\right) \eps^{\am\bm} \eps_{ab} S^b_\bm \label{Qalg1}\\
\left[J,S^a_\am\right] &=& -\frac{i}{4} a E S^a_\am + \frac{i}{4} \left( 1+ \Kin\right)  \eps^{ab} \eps_{\am\bm} Q^\bm_b. \label{Qalg2}\eea
One may then define 
\be \tQ^{\am}_a=K^{-\quarter} Q^\am_a , \quad \tS^a_\am = K^{+\quarter} S^a_\am,\ee
and in terms of these new generators the brackets are
\bea \{\tQ^\am_a, \tQ^\bm_b \} &=& -2 g \eps^{\am \bm} \eps_{ab} \sinh\left(\frac{iP}{2}\right) \label{qs1}\\
     \{\tS^a_\am, \tS^b_\bm \} &=& +2 g \eps^{ab} \eps_{\am \bm} \sinh\left(\frac{iP}{2}\right) \\
     \{\tQ^\am_a, \tS^b_\bm \} &=& \delta^b_a L^\am{}_\bm + \delta^\am_\bm R^b{}_a 
  + \delta^b_a \delta^\am_\bm\,ag E\, \eea
and
\bea \left[ J, \tQ^\am_a \right] &=& -\halfi \cosh\left(\frac{iP}{2}\right) \eps^{\am\bm}\eps_{ab} \,\tS^b_\bm \\
     \left[ J, \tS^a_\am \right] &=& +\halfi \cosh\left(\frac{iP}{2}\right) \eps^{ab}\eps_{\am\bm} \tQ^\bm_b\,.\label{qs5}\eea
For completeness, let us also list again the remaining relations involving $E,P,J$, in the following form:
\be [J,P] = aE   \qquad  [J,aE] = -i\sinh(iP) \qquad [E,P]=0\label{EQ11},\ee
\be [\tQ^\am_a, E] = [\tS^a_\am,E] = 0, \qquad [\tQ^\am_a,P]=[\tS^a_\am,P]=0.\ee  

\subsubsection*{The Plane Wave limit, Classical supersymmetry, and $\liealg d(2,1;\alpha)$}\label{pplimit}
It is interesting at this stage to remark on the plane wave limit. This is obtained by defining
\be \widehat g = ag, \qquad \widehat P = gP = \widehat g P / a,\label{PW}\ee
and then sending $a\rightarrow 0$ while keeping the hatted quantities finite. 
From the present point of view it is therefore a particular classical (in the sense of un-deformed, $q=e^{ia}\rightarrow 1$) limit of the kinematical symmetry algebra in which the coupling $g$ is also large. As noted in section \ref{Eq11}, in this regime the relations (\ref{EQ11}) reduce to the usual Poincar\'e algebra, with generators $C=agE=\widehat g E$, $\widehat P$ and $J$, and so as we expect the theory becomes relativistic, with dispersion relation
\be \frac{1}{4} = C^2 - \widehat P^2.\ee 
But we are now also free to take this limit in (\ref{qs1}-\ref{qs5}). On doing so, one obtains the following relations:
\bea  \{\tQ^\am_a, \tQ^\bm_b \} &=& -i \eps^{\am \bm} \eps_{ab} \widehat P \qquad
     \{\tS^a_\am, \tS^b_\bm \} = + i \eps^{ab} \eps_{\am \bm} \widehat P \label{cl1}\\
     \{\tQ^\am_a, \tS^b_\bm \} &=& \delta^b_a L^\am{}_\bm + \delta^\am_\bm R^b{}_a 
  + \delta^b_a \delta^\am_\bm\,C\, \eea
\be \left[ J, \tQ^\am_a \right] = -\halfi \eps^{\am\bm}\eps_{ab} \tS^b_\bm \qquad
     \left[ J, \tS^a_\am \right] = +\halfi \eps^{ab}\eps_{\am\bm} \tQ^\bm_b\,\label{cl3}\ee
\be [ J, \widehat P] = C \qquad \left[ J, C \right] = \widehat P\ee 
\be [\tQ^\am_a, C] = [\tS^a_\am,C] = 0, \qquad [\tQ^\am_a,\widehat P]=[\tS^a_\am,\widehat P]=0.\label{cl4}\ee  
These define a classical supersymmetry algebra in $1+1$ dimensions. It is unusual \cite{Freund} among super-Poincar\'e (as opposed to superconformal) algebras in that the non-abelian internal symmetries appear in the $\{\tQ,\tS\}$ bracket. The fact that this is possible is linked to the existence of the exceptional simple Lie superalgebra $\liealg d(2,1;\alpha)$.\footnote{$\liealg d(2,1;\alpha)$ was classified as a superconformal algebra in $1+1$ dimensions in \cite{GST,Gunaydin91}. See also \cite{Dictionary}.} 
Let us write the generators of the even subalgebra \be\liealg{su}(2)\times\liealg{su}(2)\times\liealg{su}(2)\subset \liealg d(2,1;\alpha)\ee as $R^a{}_b$, $L^\am{}_\bm$ and $T^{\mf a}{}_{\mf b}$, and the odd generators, transforming in the $(\bf 2,2,2)$ representation, as $F^{a\am \mf a}$. The latter close according to
\be \{F^{a\am\mf a}, F^{b\bm\mf b} \} = \alpha \eps^{bc} \eps^{\am \bm} \eps^{\mf{ab}} R^a{}_c
                                               +\beta  \eps^{ab} \eps^{\bm\cm} \eps^{\mf{ab}} L^\am{}_\cm 
                                               +\gamma \eps^{ab} \eps^{\am\bm} \eps^{\mf{bc}} T^{\mf a}{}_{\mf c} \ee
where $\alpha + \beta + \gamma = 0$. If we let $\alpha = -1- \frac{1}{R}$, $\beta=1$ and $\gamma=\frac{1}{R}$, and define
\be \widehat P =\frac{i}{2R}( T^{1}{}_{2}+T^2{}_1), \quad J = \frac{i}{2} (T^{2}{}_{1}-T^1{}_2), \quad C = \frac{1}{R} T^{1}{}_{1}=-\frac{1}{R}T^2{}_2,\ee \be \tQ^\am_a = \eps_{ab} F^{b\am 1}, \quad \tS^a_\am = \eps_{\am \bm} F^{a \bm 2} \ee
then the algebra above is recovered in the limit $R\rightarrow \8$.
This contraction is similar to that used in \cite{BeisertSU22} to obtain $\liealg{su}(2|2)\ltimes \mathbb R^2$. 

\subsubsection*{Dynamic representations via boosts}
Having seen how the supersymmetries transform under boosts, in this section we discuss the way in which the dynamic magnon representations may be obtained by boosting.

It is helpful to begin with an extremely simple example, which is sufficient to illustrate the idea. Consider the relativistic SUSY algebra $\{Q_i, Q_j\}= \left(\mathcal C\gamma^\mu\right)_{ij} P_\mu$, $i=1,2.$ There is an off-shell (``long'') multiplet  consisting of bosonic scalars $\phi, A$ and a fermionic spinor $\psi_i$. The action is $Q_i \ket\phi = \ket{\psi_i}$, $Q_i \ket{\psi_j} = \half \mathcal C_{ij} \ket A + \half \left(\mathcal C\gamma^\mu\right)_{ij} P_\mu \ket\phi$, $Q_i \ket A = P_\mu \gamma^\mu{}^j{}_i \ket{\psi_j}$. An on-shell (``short'') multiplet is obtained by setting the auxiliary $A$ to zero and requiring $\psi_i$ to obey the Dirac equation. If we take $\gamma^0 = -\sigma_3$, $\gamma^1=-i\sigma_1$, $\eta^{\mu\nu} = (+-)$, $\mathcal C=i\sigma_2$ and write $Q=Q_1$, $S=Q_2$, then 
\be \{Q,Q\}=-iP,\quad \{S,S\}=iP,\quad \{Q,S\}=E,\label{sS}\ee
a similar form to the plane-wave SUSY algebra (\ref{cl1}-\ref{cl4}). In the rest frame $\rest P_\mu= (\rest E,\rest P) = (m, 0)$, the Dirac equation $\left(\gamma\cdot P + m\right)\ket{\psi} = 0$ is $\ket{\psi_2}=0$ and the short representation is
\bea \rest Q \ket\phi = \ket{\psi_1}, &\qquad& \rest Q\ket{\psi_1} = 0\\
     \rest S \ket\phi = 0, &\qquad& \rest S\ket{\psi_1} = m \ket{\phi} .\eea
However, an observer moving with rapidity $\theta$ with respect to this frame would say that the algebra generators were $\mov O(\theta)= U \rest O U^{-1}$, $U= \exp(i\theta J)$, and would therefore see a particle with energy-momentum $P_\mu(\theta)=(E(\theta),P(\theta))= (m\cosh \theta, m\sinh \theta)$, which, since
\be Q(\theta) = \cosh(\theta/2) \rest Q -i \sinh (\theta/2) \rest S,\qquad S(\theta)= \cosh(\theta/2) \rest S +i  \sinh(\theta/2) \rest Q,\ee
carries the representation
\bea Q(\theta) \ket\phi = \cosh (\theta/2) \ket{\psi_1}, &\qquad& Q(\theta) \ket{\psi_1} = -im\sinh (\theta/2) \ket\phi\\
     S(\theta) \ket\phi = i \sinh (\theta/2) \ket{\psi_1}, &\qquad& S(\theta) \ket{\psi_1} = m\cosh (\theta/2) \ket\phi.\eea
The merit of this slightly unusual point of view, in which we choose to boost the observer rather than the particle, is that it makes it clear that the essential features of dynamic representations are present even in this simple case: the particle possesses a degree of freedom in some graded vector space, and the supersymmetries act on this space in a way that depends on the rapidity of the particle in the observer's frame.

Our claim is that each member of the 1-parameter family of dynamic magnon representations can be obtained by boosting from a rest-frame representation, just as above, with the only difference being that (\ref{sS}) is replaced by the $q$-deformed supersymmetry algebra (\ref{qs1}-\ref{EQ11}). 

We will check that this is true in the case of the fundamental multiplet $\bf (2|2)$ -- the argument for other short multiplets of $\liealg{su}(2|2)$, corresponding to magnon bound-states \cite{NDorey,CDO}, should be similar. The transformation rules for a magnon at rest ($\rest P=0$, $\rest C=ag\rest E=\half$) are
\bea \rest Q^\am_a \ket{\phi^b} = \delta^b_a \ket{\psi^\am}, &\qquad& \rest Q^\am_a \ket{\psi^\bm} = 0 \\
     \rest S^a_\am \ket{\phi^b} = 0        , &\qquad& \rest S^a_\am \ket{\psi^\bm} = \delta^\bm_\am \ket{\phi^a},\eea
while for a magnon with momentum $K(z)=e^{i\mov P(z)}$ and energy $E(z)$, both functions of the generalized rapidity $z$ as in section \ref{Eq11}, the transformation rules are
\bea Q^\am_a(z) \ket{\phi^b} = \as \delta^b_a \ket{\psi^\am}, &\qquad& 
     Q^\am_a(z) \ket{\psi^\bm} = \bs \eps^{\am\bm} \eps_{ab} \ket{\phi^b} \\
     S^a_\am(z) \ket{\phi^b} = \cs \eps^{ab} \eps_{\am \bm} \ket{\psi^\bm}, &\qquad& 
     S^a_\am(z) \ket{\psi^\bm} = \ds \delta^\bm_\am \ket{\phi^a}, \eea
or, equivalently,
\be Q^\am_a(z) = \as \rest Q^\am_a + \bs \eps^{\am\bm} \eps_{ab} \rest S^b_\bm \label{ab}\ee
\be S^a_\am(z) = \cs \eps^{ab} \eps_{\am \bm} \rest Q^\bm_b + \ds \rest S^a_\am,\label{cd}\ee
where the functions $\as,\bs,\cs,\ds$ are known \cite{AFZ,Beisert06} to be
\bea \as &=&e^{i\xi} \sqro{agE(z)+\alf} \label{ab1} \\ \bs &=& e^{-i\xi} \sqri{agE(z)+\alf} g(1-K(z)) \\
     \cs &=& e^{i\xi}\sqri{agE(z)+\alf} g(1-K(z)^{-1})\label{cd1} \\ \ds &=& e^{-i\xi} \sqro{agE(z)+\alf}.\eea
We are using here the ``string theory basis'' of \cite{AFZ}, i.e. $\eta\sim \sqrt \zeta$ in the notation of that paper,\footnote{In terms of the variables $x^\pm$ \cite{Beisert06}, $K=\frac{x^+}{x^-}$, $C=agE=-igx^++igx^- -\alf$ $\Rightarrow$ $x^-=\frac{agE+\half}{ig(1-K)}$, $x^+=Kx^-$.} and $\xi$ is a phase that may in principle depend on the rapidity.

Now the claim is that, with $\mov P(z) = e^{i z J} \rest P e^{-izJ}$ and $\mov E(z) = e^{i z J} \rest E e^{-izJ}$, 
\be Q^\am_a(z) = e^{i z J} \rest Q^\am_a e^{-izJ}, \quad S^a_\am(z) = e^{i z J} \rest S^a_\am e^{-izJ}.\label{clm}\ee
Assuming $e^{i\xi(0)}= 1$, this certainly true when $z=0$, so it suffices to check that the $z$-derivatives of both sides are equal. Using (\ref{dr2}) and (\ref{Qalg1}-\ref{Qalg2}), one finds the following differential equations on combining (\ref{ab}-\ref{cd}) and (\ref{clm}):\\
\be \frac{\del\as}{\del z} = -\frac{1}{4} aE(z)\as + \frac{1}{4}(1+K(z)) \cs \quad\Rightarrow\quad
    \frac{\del}{\del z} (\as K(z)^{-\quarter}) = \frac{1}{4}(1+K(z)) \cs K(z)^{-\quarter},\ee
\be \frac{\del\cs}{\del z} = \frac{1}{4} aE(z)\cs - \frac{1}{4}(1+K(z)^{-1}) \as \quad \Rightarrow \quad
    \frac{\del}{\del z} (\cs K(z)^\quarter) = - \frac{1}{4}(1+K(z)^{-1}) \as K(z)^\quarter,\ee
together with similar equations relating $\bs$ and $\ds$. The functions in (\ref{ab1}) and (\ref{cd1}) do indeed satisfy these, provided we take 
\be e^{i\xi(z)} = K^\quarter(z)\quad\Leftarrow\quad \xi(z)=\quarter P(z),\ee
and we are done. (In verifying this, a useful form of the on-shell condition (\ref{casval}) turns out to be $(K-K^{-1})(1-K^{-1}) = \frac{1}{g^2}(1+K^{-1})(\alf+agE)(\alf-agE)$.) 

The procedure above should be compared with the result \cite{Beisert06} that $\liealg{su}(2|2)\ltimes \mathbb{R}^2$ possesses an $\liealg{sl}(2)$ of outer automorphisms that can be used to rotate the triplet of central charges $(\Pp,\Pm,C)$ into the form $(0,0,C')$, which allows the representations of $\liealg{su}(2|2)\ltimes \mathbb{R}^2$ to be placed in correspondence with those of $\liealg{su}(2|2)$. The main difference is that, there, the transformations were classical and in an internal space, whereas here, at the cost of being $q$-deformed, they have the status of spacetime symmetries of the two dimensional scattering theory.

\section{Coalgebra structure}\label{coalgebra}
So far everything we have said concerns the kinematical symmetries of individual excitations. The natural question is whether the deformed super-Poincar\'e algebra in (\ref{qs1}-\ref{EQ11}) is also a dynamical symmetry -- which is to ask, in the context of the present system, whether it is a symmetry of the $S$-matrix. A full answer to this question is beyond the scope of the present work, but in this section we at least outline what needs to be done and comment on a couple of interesting features.

To say that the two-particle $S$-matrix possesses a given symmetry $X$ means that
\be \left[ S, \Delta X\right] = 0.\ee
Here the coproduct specifies how $X$ acts on pairs of excitations. (If particles $\chi_1$ and $\chi_2$ transform in representations $\pi_1$ and $\pi_2$ then the tensor product state $\chi_1 \otimes \chi_2$ transforms in the representation $(\pi_1\otimes \pi_2) \circ \Delta $.) Certainly then, before even looking at $S$, the first thing one needs is the correct coalgebra structure. This is not trivial to find, and indeed we have not shown that any consistent coproduct exists for (\ref{qs1}-\ref{EQ11}) as they stand.
Here we focus on the subalgebra of (\ref{qs1}-\ref{EQ11}) generated by $E,P,J$ and the two particular supersymmetries
\be \tQ = \frac{1}{\sqrt 2}\left(\tQ^1{}_2 +\tQ^2{}_1\right), \qquad 
    \tS = \frac{1}{\sqrt 2}\left(\tS^1{}_2 + \tS^2{}_1\right).\ee
These obey
\bea \{ \tQ, \tQ\} = -2g \sinh \left(\frac{iP}{2}\right),& \qquad& \left[ J, \tQ \right] = -\frac{i}{2} \cosh\left(\frac{iP}{2}\right) \tS, \label{ook1}\\
\{ \tS, \tS \} = 2g \sinh \left(\frac{iP}{2}\right), &\qquad&  \left[ J, \tS \right] = +\frac{i}{2} \cosh\left(\frac{iP}{2}\right) \tQ, \eea
\be \{\tQ, \tS\}= agE, \qquad \left[E,P\right]=0,\ee
\be  \left[J,P\right] = aE,   \qquad  \left[J,aE\right] = -i\sinh(iP). \label{ook4}\ee
By construction, the internal symmetries $L$ and $R$ are now absent, and the even-graded subalgebra is just $E_q(1,1)$ as in section \ref{Eq11}. This comes equipped with a coalgebra structure inherited from that of $U_q(\liealg{sl}(2))$, of which $E_q(1,1)$ is a limit \cite{CGST}:
\be \Delta K = K \otimes K \quad \Leftarrow \quad \Delta P = 1\otimes P + P \otimes 1,\label{ook5}\ee
\be \Delta E = K^{-\half} \otimes E + E \otimes K^\half \label{ook6}\ee
\be \Delta J = K^{-\half} \otimes J + J \otimes K^\half.\label{DJ1}\ee  
Before proceeding, one puzzle should be noted: here $\Delta E$ is non-trivial, whereas in the spin chain the energy should act additively on multi-excitation states.

A natural guess for the coproducts of the odd-graded generators is\footnote{Given the earlier definitions $\mathcal Q^\am_a=Q^\am_a K^{-\quarter}$ and $\mathcal S^a_\am= S^a_\am K^{\quarter}$, these are consistent with the following coproducts of the original $Q$, $S$:
\be \Delta Q^\am_a = Q^\am_a \otimes K^\half + 1 \otimes Q^\am_a, \quad \Delta S^a_\am = S^a_\am \otimes 1 + K^{-\half} \otimes S^a_\am,\ee
and here $\Delta Q^\am_a$ accords with the intuition that if $\Pp$ inserts a $Z$, then we can symmetrize things so that each $Q^\am_a$ inserts ``half a $Z$''.}
\be \Delta \tQ = K^{-\quarter} \otimes \tQ + \tQ \otimes K^\quarter \label{ooko}\ee
\be \Delta \tS = K^{-\quarter} \otimes \tS + \tS \otimes K^\quarter.\label{ookoo}\ee
These are compatible with the brackets $\{\tQ,\tQ\}$, $\{\tQ,\tS\}$, $\{\tS,\tS\}$, in the sense that $\Delta$ is a homomorphism of algebras -- $\{\Delta \tQ,\Delta \tQ\} = -2g\sinh (i\Delta P/2)$ and so on -- but \emph{not} with the brackets $[J,\tQ]$ and $[J,\tS]$. However, some experimentation reveals that it is possible to modify the coproduct (\ref{DJ1}) of $J$ in a way which fixes this, without spoiling any other relations:
\be \Delta J = K^{-\half} \otimes J + J \otimes K^\half - \frac{i}{4g} K^{-\quarter} \left( \tQ\otimes \tS + \tS\otimes \tQ\right) K^\quarter.\label{DJ2}\ee
We thus have a graded bialgebra -- call it $\mathcal B$ -- generated by $E,P,J,\tQ,\tS$, defined by the algebra relations (\ref{ook1}-\ref{ook4}) and the coalgebra relations (\ref{ook5}-\ref{ook6}), (\ref{ooko}-\ref{ookoo}) and (\ref{DJ2}). It has previously in been obtained \cite{LN} as a contraction limit of the simple $q$-deformed superalgebra $U_q(\liealg{osp}(1|2))$. The point to note is that although the deformed Poincar\'e algebra $E_q(1,1)$ is contained in $\mathcal B$, and closes as an algebra, it is \emph{not} closed under the coproduct within $\mathcal B$. 

An important consequence is that, whereas boost-covariance of an $S$-matrix usually manifests itself as a differential equation $(\del_{\theta_1} + \del_{\theta_2})S_{ij}(\theta_1,\theta_2)=0$ satisfied by each matrix element individually, here one should not expect this to be true. Whenever terms like $\tQ\otimes \tS$ occur in $\Delta J$, boost-covariance, if present, will take the form of some more complicated matrix equation satisfied by $S$. 

The conclusion to draw is that one needs to find the correct bialgebra structure, and most importantly the full form of the coproduct of $J$. To attack this problem, it would be interesting to focus on limiting cases such as the giant magnon \cite{GM} or near-plane-wave regimes, and to make contact with the results of \cite{KMRZ} (see also \cite{AFPZ}) on world-sheet scattering in the gauge-fixed $AdS_5 \times S^5$ sigma-model. Finally, insight into the coproduct here might be gained from comparison with the Yangian of $\liealg{su}(2|2)$ which appeared recently in \cite{B07}.

\vspace{1.5cm} 
{\bf Acknowledgements} I have benefited from useful discussions with Nicolas Cramp\'e, Patrick Dorey, David Kagan and Marija Zamaklar, and from a helpful exchange of emails with C\'esar G\'omez and Rafael Hern\'andez. I gratefully acknowledge the financial support of the Leverhulme Trust.

\end{document}